\documentclass[sigplan]{acmart}

\acmConference[]{A Diagrammatic Calculus for Algebraic Effects}
\acmISBN{} 
\acmDOI{} 
\startPage{1}

\setcopyright{none}

\usepackage{booktabs}   
\usepackage{subcaption} 

\usepackage{graphicx}
\usepackage{multicol}
\usepackage{proof}
\usepackage[all]{xy}

\usepackage{framed}
\usepackage{floatpag}
\usepackage{mdframed}
\usepackage{sansmath}
\usepackage{mathtools}
\usepackage[bbgreekl]{mathbbol}
\usepackage{graphbox}

\usepackage{amssymb}
\usepackage{color}
\usepackage{url}
\usepackage{stmaryrd}
\usepackage{xypic}

\newcounter{numberone}
\newenvironment{varenumerate}
{
\begin{list}{\arabic{numberone}.}
{
  \usecounter{numberone}
  \setlength{\itemsep}{0pt}
  \setlength{\topsep}{0pt}
  \setlength{\parsep}{0pt}
  \setlength{\partopsep}{0pt}
  \setlength{\leftmargin}{15pt}
  \setlength{\rightmargin}{0pt}
  \setlength{\itemindent}{0pt}
  \setlength{\labelsep}{5pt}
  \setlength{\labelwidth}{15pt}
}}
{
\end{list}
} 

\theoremstyle{definition} 
\newtheorem{remark}{Remark}

\newcommand{\hide}[1]{}
\RequirePackage{ifthen}
\newcommand{\typeof}{0}
\newcommand{\longv}[1]{\ifthenelse{\equal{\typeof}{0}}{}{#1}}
\newcommand{\shortv}[1]{\ifthenelse{\equal{\typeof}{0}}{#1}{}}

\newcommand{\values}{\mathcal{V}}

\newcommand{\subst}[3]{#1[#2:=#3]}

\newcommand{\valone} {V}
\newcommand{\varone} {x}
\newcommand{\vartwo} {y}

\newcommand{\termone}{e}
\newcommand{\termtwo}{f}
\newcommand{\termthree}{g}

\newcommand{\support}[1]{\text{\emph{supp}}(#1)}

\newcommand{\signature}{\Sigma}

\newcommand{\abs}[1]{\lambda #1.}

\newcommand{\lan}{\langle}
\newcommand{\ran}{\rangle}
\newcommand{\cc}{\cdots}
\newcommand{\hh}{\hdots}

\newcommand{\set}{\mathsf{Set}}
\newcommand{\powerset}{\mathcal{P}}
\newcommand{\distribution}{\mathcal{D}}

\newcommand{\inl}{\mathsf{in}_l}
\newcommand{\inr}{\mathsf{in}_r}
\newcommand{\monad}{T}
\newcommand{\unit}{\eta}

\newcommand{\defeq}{\triangleq}

\newcommand{\cpoleq}{\sqsubseteq}

\newcommand{\sem}[1]{\llbracket #1 \rrbracket}

\newcommand{\kleisli}[1]{#1^{\dagger}}

\renewcommand{\valone}{v}
\newcommand{\valtwo}{w}

\renewcommand{\termone}{t}
\renewcommand{\termtwo}{s}
\renewcommand{\termthree}{r}

\newcommand{\TO}[1]{\Rightarrow^{n}}

\newcommand{\print}[2]{\mathbf{print}_{#1}. #2}

\newcommand{\comp}{\mathbin{\circ}}

\renewcommand{\monad}{T}

\renewcommand{\set}{\mathbf{Set}}
\renewcommand{\powerset}{\mathcal{P}}

\newcommand{\op}{\textnormal{\texttt{op}}}

\newcommand{\seq}[2]{\mathbf{let}\ \varone = #1\ \mathbf{in}\ #2}

\newcommand{\seqy}[2]{\mathbf{let}\ \vartwo = #1\ \mathbf{in}\ #2}

\renewcommand{\inl}[1]{\mathbf{inl}\ #1}
\renewcommand{\inr}[1]{\mathbf{inr}\ #1}

\renewcommand{\powerset}{\mathcal{P}}
\newcommand{\divergence}{\uparrow}

\newcommand{\bindsymbol}{\scalebox{0.5}[1]{$>\!>=$}}
\newcommand{\bind}{\mathrel{\bindsymbol}}

\newcommand{\maybe}{\mathcal{M}}

\newcommand{\exception}{\mathcal{E}}

\renewcommand{\distribution}{D}

\newcommand{\globalstate}{\mathcal{G}}

\newcommand{\Output}{\mathcal{O}}

\newcommand{\printsymbol}[1]{\mathbf{print}_{#1}}

\makeatletter
\newcommand{\uset}[3][0ex]{%
  \mathrel{\mathop{#3}\limits_{
    \vbox to#1{\kern-7\ex@
    \hbox{$\scriptstyle#2$}\vss}}}}
\makeatother

\newcommand{\eff}[4]{
\boxed{#1}
\raisebox{0.4em}{$\underbar{$\overset{\scriptstyle\makeset{#2}}{\quad}$ }$}
\raisebox{-0.439em}{$\overline{ \scriptstyle #3\quad  }$} #4
}

\newcommand{\effm}[4]{
\boxed{#1}
\raisebox{0.4em}{$\underbar{$\overset{\scriptstyle\hspace{0.1cm}\makeset{#2}}{\quad}$ }$}
\raisebox{-0.439em}{$\overline{ \scriptstyle #3\quad  }$} #4
}

\newcommand{\efftwo}[7]{
\boxed{#1}
\raisebox{0.4em}{$\underbar{$\overset{\scriptstyle\makeset{#2}}{\quad}$ }$}
\raisebox{-0.439em}{$\overline{ \scriptstyle #3\quad  }$}
\boxed{#4}
\raisebox{0.4em}{$\underbar{$\overset{\scriptstyle\hspace{0.1cm}\makeset{#5}}{\quad}$ }$}
\raisebox{-0.439em}{$\overline{ \scriptstyle #6\quad  }$} #7
}

\newcommand{\effnull}[2]
{
\boxed{#1}
\raisebox{0.4em}{$\underbar{$\overset{\scriptstyle\phantom{n}}{\quad}$ }$}
\raisebox{-0.439em}{$\overline{ \scriptstyle \phantom{i}\quad  }$} #2
}

\newcommand{\effk}[4]{
\boxed{#1}
\raisebox{0.4em}{$\underbar{$\overset{\scriptstyle[#2]}{\quad}$ }$}
\raisebox{-0.51em}{$\overline{ \scriptstyle #3\quad  }$} #4
}

\newcommand{\effunit}[1]{
  \boxed{H}
\raisebox{0.4em}{$\underbar{$\overset{\scriptstyle\phantom{[1]}}{\quad}$ }$}
\raisebox{-0.439em}{$\overline{ \scriptstyle \phantom{1}\quad  }$} #1
}

\newcommand{\effthree}[9]{
  \boxed{#1}
\raisebox{0.4em}{$\underbar{$\overset{\scriptstyle\makeset{#2}}{\quad}$ }$}
\raisebox{-0.439em}{$\overline{ \scriptstyle #3\quad  }$}
\boxed{#4}
\raisebox{0.4em}{$\underbar{$\overset{\scriptstyle\hspace{0.1cm}\makeset{#5}}{\quad}$ }$}
\raisebox{-0.439em}{$\overline{ \scriptstyle #6\quad  }$} 
\boxed{#7}
\raisebox{0.4em}{$\underbar{$\overset{\scriptstyle\hspace{0.1cm}\makeset{#8}}{\quad}$ }$}
\raisebox{-0.51em}{$\overline{ \scriptstyle #9\quad  }$}
}

\newcommand{\effconvergence}[1]{
 \boxed{\downarrow}
\raisebox{0.4em}{$\underbar{$\overset{\phantom{\scriptstyle[1]}}{\quad}$ }$}
\raisebox{-0.439em}{$\overline{\phantom{\scriptstyle{1}}\quad}$} #1
}
\newcommand{\boteff}{
  \boxed{\bot}
}

\newcommand{\makeset}[1]{\boldsymbol{#1}}

\newcommand{\mtermone}{\xi}
\newcommand{\mtermtwo}{\rho}
\newcommand{\mtermthree}{\nu}

\newcommand{\vect}[1]{\vec{#1}}

\renewcommand{\varone}{x}
\renewcommand{\vartwo}{y}
\renewcommand{\valone}{v}
\renewcommand{\valtwo}{w}

\renewcommand{\maybe}{\mathsf{M}}

\renewcommand{\distribution}{\mathcal{D}}
\renewcommand{\powerset}{\mathcal{P}}

\newcommand{\gop}{\Gamma}
\newcommand{\goptwo}{\Delta}
\newcommand{\gopthree}{\Xi}

\renewcommand{\op}{\sigma}

\begin{document}

\title{A Diagrammatic Calculus for Algebraic Effects}        
\subtitle{Preliminary Report}                    

\author{Ugo Dal Lago}
\affiliation{
\institution{University of Bologna and INRIA Sophia Antipolis}           
}
\email{ugo.dallago@unibo.it}         

\author{Francesco Gavazzo}
\affiliation{
\institution{University of Bologna and INRIA Sophia Antipolis}         
}
\email{francesco.gavazzo2@unibo.it}        

\begin{abstract}
We introduce a new diagrammatic notation for representing the result 
of (algebraic) effectful computations. Our notation explicitly separates 
the effects produced during a computation from the possible values returned, 
this way simplifying the extension of definitions and results on pure computations 
to an effectful setting. Additionally, we show a number of algebraic and 
order-theoretic laws on diagrams, this way laying the foundations for 
a diagrammatic calculus of algebraic effects.
We give a formal foundation for such a calculus in terms of Lawvere theories and 
generic effects.
\end{abstract}

\maketitle

\section{Introduction}
\label{sect:introduction}

In this note we are concerned with the problem of finding a convenient presentation 
of effectful (sequential) computations when effects are produced by 
\emph{algebraic operations} \cite{PlotkinPower/FOSSACS/02,Plotkin/algebraic-operations-and-generic-effects/2003}.
Concretely, what we have in mind are
computations generated by (sorts of) 
$\lambda$-calculi enriched with algebraic operations in the 
spirit of Plotkin and Power \cite{PlotkinPower/FOSSACS/01}. 
Say we have fixed such a $\lambda$-calculus: 
how do we represent the result $\sem{\termone}$ of the evaluation of a 
term $\termone$? Two standard answers to this question are the following:
\begin{enumerate}
    \item $\sem{\termone}$ is a possibly infinite tree whose nodes are labeled with 
        algebraic operations, and whose leaves are either values or the bottom 
        symbol $\bot$ denoting pure divergence. These structures are known as 
        computation trees \cite{Goguen/1977} and are used to give operational semantics 
        to calculi with algebraic effects in, e.g., \cite{PlotkinPower/FOSSACS/01,Simpson-Niels/Modalities/2018,JohannSimpsonVoigtlander/LICS/2010}. 
    \item $\sem{\termone}$ is an element in $\monad(\values)$, where 
        $\monad$ is a suitable monad and $\values$ is the set of values. 
        This is the approach followed in, e.g., \cite{DalLagoGavazzoLevy/LICS/2017,normalFormBisimulation}. 
\end{enumerate}

Both these approaches have their drawback. The first one is essentially syntactical, 
and does not give any semantical information on the effects produced. The second, on 
the contrary, is purely semantical and does not allow one to separate terms/values from 
effects. This is rather unsatisfactory as effects are performed by operations, and thus 
we would expect $\sem{\termone}$ to be made of an `effect part' and a 
`value part', 
rather than being a monolithic object. 

In this note we show that we can rely on the correspondence between algebraic effects and 
\emph{generic operations} \cite{Plotkin/algebraic-operations-and-generic-effects/2003} 
to express $\sem{\termone}$ as a pair $(\gop, \vect{\valone})$, 
where $\gop$ is a (generic) 
effect and $\vect{\valone}$ is a list of values. 
In order to facilitate calculations 
with generic effects we present the pair $(\gop, \vect{\valone})$ diagrammatically, 
this way obtaining a lightweight diagrammatic calculus for effectful sequential 
computations.

\paragraph{Advertisement} 
This note is a work in progress. The authors plan to 
systematically review and update it---especially with examples and informal explanations---in the next weeks. The authors have also noticed that 
different browsers have different rendering of diagrams, whereas `desktop' 
pdf viewers tend to have a better rendering.

\section{Preliminaries: Monads and Algebraic Operations} 
\label{sect:monads}
In this section, we give some background notions on monads and algebraic operations. 
The current version of this work still lacks a proper introduction to monads and 
algebraic effects, and thus assume the reader to be familiar with basic category theory 
\cite{MacLane/Book/1971} and
domain theory \cite{AbramskyJung/DomainTheory/1994}. 

We work with (strong) monads\footnote{Recall that any monad on 
$\set$ is strong.} on $\set$, the category of set and functions, 
which we present as 
\emph{Kleisli triples} \cite{MacLane/Book/1971}. The latter are triples 
$(\monad, \unit, \kleisli{-})$, where $\kleisli{f}: \monad X \to 
\monad Y$ is the Kleisli extension of $f: X \to \monad Y$
and $\unit: X \to \monad X$ is the unit of $\monad$, satisfying the following laws, 
for $f,g$ having appropriate (co)domains.
\begin{align*}
\kleisli{\unit_X} &= id_{\monad(X)};
&
\kleisli{f} \comp \unit_X &= f;
&
\kleisli{g} \comp \kleisli{f}
&= \kleisli{(\kleisli{g} \comp f)}.
\end{align*}

Since we work in $\set$, we oftentimes use the bind notation $\bindsymbol$ 
for monadic sequencing. That is, given $\mu \in \monad(X)$ and $f: X \to \monad(Y)$, 
we write $\mu \bind f$ in place of $\kleisli{f}(\mu)$.

We assume $\unit$ to be an injection, meaning that
$\monad$ is \emph{non-trivial} \cite{Manes/Taut-monads}. 
This is the case for  
all monads on $\set$, except for the monad with 
$\monad(X) = 1$ for every set $X$, and the one with 
$\monad(\emptyset) = \emptyset$ and $\monad(X) = 1$, for $X \neq \emptyset$.
 We denote by $Kl(\monad)$ the Kleisli category of $\monad$.

\begin{example} 
\label{ex:monads}
The target monads we have in mind are those modeling notions of computations 
\cite{Moggi/LICS/89,Moggi/Notions-Of-Computations/1991,DBLP:conf/afp/Wadler95}. 
Among those are: 
\begin{varenumerate}
    \item The maybe monad $\maybe(X) \defeq X + \{\divergence\}$ modeling divergence.
    \item The exception monad $\exception(X) \defeq X + E$ 
      modeling computations raising exceptions in a given set $E$.
      from a given  
    \item The powerset monad $\powerset$ modeling pure nondeterminism.
    \item The (discrete) subdistribution monad $\distribution$ modeling probabilistic 
      nondeterminism.
    \item The global state monad $\globalstate(X) \defeq \maybe(X \times S)^S$, 
      modeling imperative computations over a store $S$ (e.g. given a set $L$ of
      locations, take $S \defeq \{0,1\}^L$, meaning that a store assigns to each 
      location a boolean value).
    \item The output monad $\Output(X) \defeq \mathcal{C}^\infty \times \maybe(X)$, 
      where $\mathcal{C}^{\infty}$ is the set of finite and infinite strings 
      over a fixed alphabet $\mathcal{C}$.
    \item Combinations thereof, such as monads of the form $\monad(\maybe(X))$ or 
      $\monad(X \times S)^S$ (see \cite{DBLP:journals/tcs/HylandPP06}).
\end{varenumerate}
\end{example}

Monads alone can structure effects and (sequentially) compose them, but 
do not have the structure to actually produce them (with the irrelevant exception 
of the trivial effect). For this reason we consider monads coming with 
effect-triggering operations.
Following Plotkin and Power \cite{DBLP:journals/entcs/PlotkinP01,PlotkinPower/FOSSACS/01},
we require such operations to be algebraic.

\begin{definition}
\label{def:algebraic-operation}
An $n$-ary (set-indexed family of) operation
$\op_X: \monad X^n \to \monad X$ is \emph{algebraic}, 
if for all sets $X,Y$, $f: X \to \monad(Y)$, and $\mu_1, \hh, \mu_n \in \monad(X)$,  
we have:
\begin{align*}
(\op_X(\mu_1, \hh, \mu_n)) \bind f = \op_Y(\mu_1 \bind f, \hh, \mu_n \bind f).
\end{align*}
\end{definition}

\begin{example}
\label{ex:algebraic-operations}
Referring to the monads in Example~\ref{ex:monads}, 
the following operations are all algebraic.
\begin{varenumerate}
  \item The maybe monad comes with no operation.
  \item The exception monad comes with a set of zero-ary operations 
    $\mathbf{raise}_e$, for $e \in E$, that simply corresponds to 
    $\inr(e)$. 
  \item The powerset monad $\powerset$ comes with set-theoretic union $\cup$ 
    modeling binary nondeterministic choice.
    \item The subdistribution monad $\distribution$ comes with binary fair 
      nondeterministic choice $\oplus$ (recall that $(\mu \oplus \nu)(x) = 
      \frac{1}{2} \mu(x) + \frac{1}{2} \nu(x)$).
    \item The global state monad $\globalstate(X) \defeq \maybe(X \times S)^S$, 
      comes with a set of binary operations $\mathbf{read}_{\ell}$ for reading locations,
       and a set of unary operations $\mathbf{write}_{\ell}$ for writing locations 
       (see \cite{DBLP:journals/entcs/PlotkinP01,Pretnar/Tutorial-algebraic-effects/2015}).
    \item The output monad comes with a set of unary operations $\printsymbol{\mathtt{c}}$ 
      indexed over elements of $\mathcal{C}$ for printing. That is, 
      $\print{\mathtt{c}}{(w,x)} = (cw, x)$.
\end{varenumerate}
\end{example}

As highlighted by Example\ref{ex:algebraic-operations} several computational 
effects can be modeled using monads and algebraic operations. Notable exceptions 
are continuations and exception handling. In this paper we will consider 
computational effects modeled as a monad $\monad$ together with a set $\signature$ of 
algebraic operations on $\monad$. In those cases, we say that $\monad$ is 
$\signature$-algebraic.

\section{A $\lambda$-Calculus with Algebraic Effects}
At this point of the work we have to make some design choices, as well as 
introduce some (minor) restrictions on the collection of monads and effects 
studied. In order to motivate such choices and restrictions, it is convenient 
to have a concrete computational calculus with algebraic effects. 
We take the calculus of \cite{DalLagoGavazzoLevy/LICS/2017}.

\begin{definition}
\label{def:lambda-sigma}
Let $\signature$ be a set of operations on a given monad $\monad$. 
The calculus $\Lambda_{\signature}$ has terms and values defined by the following 
grammar, where $\varone$ ranges over a countable set of variables, and 
$\op \in \signature$.
\begin{align*}
\termone, \termtwo &::= \varone \mid \abs{\varone}{\termone} \mid \termone\termtwo \mid 
  \op(\termone, \hh, \termone)
\\
\valone,\valtwo &::= \varone \mid \abs{\varone}{\termone}
\end{align*}
\end{definition}

Notice that $\Lambda_{\signature}$ is parametric with respect to a set $\signature$ 
of operations and to a $\signature$-algebraic monad $\monad$. Although the monad 
$\monad$ plays no role in Definition~\ref{def:lambda-sigma} and calculi are usually defined 
relying on uninterpreted operation symbols (which are then interpreted as algebraic 
operations once giving semantics to the calculus), for the sake of the economy 
of the work we chose the `semantic-oriented' presentation of 
Definition~\ref{def:lambda-sigma}. 

\begin{example}
\begin{varenumerate}
   \item Taking the maybe monad (and no operation) 
      we obtain the pure $\lambda$-calculus.
    \item Taking the exception monad with the raising exception operation(s) 
      of Example~\ref{ex:algebraic-operations} we obtain a $\lambda$-calculus 
      with exceptions.
    \item Taking the powerset with set-theoretic union (as 
      in Example~\ref{ex:algebraic-operations}), we obtain the nondeterminism 
      $\lambda$-calculus \cite{Ong/LICS/1993,deLiguoroPiperno/IC/1995,Lassen/PhDThesis}.
    \item Taking the subdistribution monad $\distribution$ 
      and fair probabilistic choice (as in Example~\ref{ex:algebraic-operations}) 
      we obtain the probabilistic $\lambda$-calculus of 
      \cite{DalLagoZorzi/TIA/2012}.
    \item Taking the global state monad and the operation for reading and 
      writing stores of Example~\ref{ex:algebraic-operations}
      we obtain the imperative $\lambda$-calculus.
    \item Taking the output monad and the printing operation(s) of 
      Example~\ref{ex:algebraic-operations} we obtain the $\lambda$-calculus 
      with output \cite{DBLP:phd/ethos/Gordon92,10.1007/3-540-61064-2_46}.
\end{varenumerate}
\end{example}
We follow standard notational conventions, as in \cite{Barendregt/Book/1984}. 
In particular, we denote by $\Lambda$ and $\values$ the collection of \emph{closed} 
terms (programs) and values.
Additionally, we write $\seq{\termone}{\termtwo}$ for $(\abs{\varone}{\termtwo})\termone$.

Next, we want to give, say call-by-value, semantics to $\Lambda_{\signature}$. 
To do so, we follow \cite{DalLagoGavazzoLevy/LICS/2017} and give a monadic 
semantics to the calculus. That is, to any closed term $\termone$ 
is associated an element $\sem{\termone}$ in 
$\monad(\values)$ (we call such elements monadic values).
Intuitively, the map $\sem{-}: \Lambda \to \monad(\values)$ should be defined 
as follows.

\begin{align*}
\sem{\valone} &\defeq \unit(\valone)
\\
\sem{\termone \termtwo} 
&\defeq \sem{\termone} \bind 
(\valone \mapsto (\sem{\termtwo} \bind (\valtwo \mapsto \sem{\valone\valtwo})))
\\
\sem{\op(\termone_1,\hh, \termone_n)} &\defeq \op(\sem{\termone_1}, \hh, \sem{\termone_n})
\end{align*}

Obviously, a map $\sem{-}$ defined in such a way may not exist. In order to ensure 
the existence of $\sem{-}$ \cite{DalLagoGavazzoLevy/LICS/2017} required $\monad$
to come with a suitable domain structure (notably, $\omega$-cppo enrichment 
\cite{Kelly/EnrichedCats}).

\begin{definition}
Let $\monad$ be a $\signature$-algebraic monad. 
We say that $\monad$ is $\signature$-\emph{continuous} if for any
set $X$, $\monad X$ carries an $\omega$-cppo structure such that both
$\bindsymbol$ and operations in $\signature$ are continuous in 
all arguments.
\end{definition}

\begin{proposition}[\cite{DalLagoGavazzoLevy/LICS/2017}]
If $\monad$ is $\signature$-continuous, $\sem{-}: \Lambda \to \monad(\values)$ 
is the \emph{least} map satisfying the following laws:
\begin{align*}
\sem{\valone} &\defeq \unit(\valone)
\\
\sem{\termone \termtwo} 
&\defeq \sem{\termone} \bind 
(\valone \mapsto (\sem{\termtwo} \bind (\valtwo \mapsto \sem{\valone\valtwo})))
\\
\sem{\op(\termone_1,\hh, \termone_n)} &\defeq \op(\sem{\termone_1}, \hh, \sem{\termone_n})
\end{align*}
\end{proposition}

\begin{example}
All monads in Example~\ref{ex:monads} are $\signature$-continuous (wrt 
the set of operations of Examples~\ref{ex:algebraic-operations}). 
The $\omega$-cppo structure is standard (e.g. the one induced by subset inclusion 
for $\powerset$, or by the pointwise order for $\distribution$), with the exception of 
the one for the output monad, which is induced by the following order:
\begin{align*}
(u,\mu)\cpoleq (w,\nu) 
&\iff 
(\mu = \inr(\divergence) \wedge u \cpoleq w) 
\\
&\quad \vee (\mu = \inl(x) = \nu \wedge u=w).
\end{align*}
\end{example}

\section{Countable Monads and Generic Effects}

In previous section we defined the (result of the) evaluation of 
a program $\termone$ as a monadic value 
$\sem{\termone} \in \monad(\values)$. Working with monadic values
has the major advantage 
of providing semantical information on the effects performed. 
However, it also has a major drawback: $\sem{\termone}$ being 
a monolithic object, it does not allow a clear distinction 
between the effects produced by $\termone$ and the possible results (values) obtained. 

Intuitively, since during the evaluation of $\termone$ effects 
can only be produced by (algebraic) operations, we would expect $\sem{\termone}$ 
to be a pair 
of the form $(\gop, \lan \valone_i \ran_{i \in I}\ran)$, where $\gop$ is a mathematical 
object describing the effects produced by $\termone$, and $\lan \valone_i \ran_{i \in I}$
is the list of possible values returned. 

Notice that since the calculus may be 
nondeterministic, $I$ has cardinality bigger than one in general, and that due to 
recursion $I$ may be infinite (but countable). To see that, consider the nondeterministic 
program $\termone \boldsymbol{0}$, where $\boldsymbol{n}$ stands for the (Church) numeral 
of $n$, $\mathbf{succ}$ is the successor function, 
and $\termone$ is recursively defined by the equation 
$\termone = \abs{x}{(x \cup \termone(\mathbf{succ}\ x))}$.

The purpose of this section is to prove that the aforementioned composition 
is indeed always possible.

\subsection{Countable Monads}

First, we have to show that any monadic value has indeed a `value component'. 
Mimicking the terminology employed for subdistributions, we refer to such a 
component as \emph{support}.
As the set of terms is countable, we expect any monadic value to have countable 
support. 
In general, not all monads come with a notion of support, and even those 
that have such a notion may have objects with uncountable support (e.g. the 
continuation monad \cite{DBLP:journals/entcs/HylandP07}). As a consequence, 
we need to isolate the class of monads whose elements have countable support.

First, we need monad to preserves injections (the reason why we need such monads 
will become clear soon). That is, if $A \xhookrightarrow{\ \iota\ } X$ is the subset inclusion map, then  
$\monad (A) \xhookrightarrow{\ \monad \iota\ } \monad (X)$ is an injection (i.e. a mono), 
which we regard as monadic inclusion. Notice that if $\monad$ preserves 
weak pullbacks, then it also preserves monos. This condition is met by all the monads 
in Example~\ref{ex:monads} (see, e.g., \cite{Jacobs/Introduction-to-coalgebra/2016}).

Given a monadic object $\mu \in \monad (X)$, the support of $\mu$ 
is the \emph{smallest} 
set $A \xhookrightarrow{\ \iota\ } X$ such that $\mu \in \monad (A)$. 
We denote such a set by $\support{\mu}$. 
Of course, in general the support of $\mu$ need not 
exist and thus we restrict our analysis to monads coming 
with a notion of \emph{countable} support.

\begin{definition}
We say that a monad is \emph{countable} if
for any set $X$ and any element $\mu \in \monad(X)$, there exists 
a smallest countable set $Y \xhookrightarrow{\ \iota\ } X$, denoted by 
$\support{\mu}$, such that 
$\mu \in \monad(Y)$ (i.e. there exists $\nu \in \monad(Y)$ such that 
$\mu = (\monad\iota)(\nu)$).
\end{definition}

\begin{example}
All the monads in Example~\ref{ex:monads} are countable, with the 
exception of the powerset monad. Nonetheless, we can
regard $\powerset$ as countable (by taking its countable restriction),
the collection of $\lambda$-terms being countable itself. 
\end{example}

\begin{remark}
The notion of support is sometimes formalized throughout the notion of an 
\emph{accessible} functor (monad, in our case). Accordingly, a monad $\monad$ 
is $\kappa$-accessible, for a cardinal $\kappa$, if for any element 
$\mu \in \monad(X)$ there exists $A \subseteq X$ with cardinality strictly 
smaller than $\kappa$ such that $\mu \in \monad(A)$. 
If $\kappa = \aleph_0$, then $\monad$ is said to be \emph{finitary} 
as, intuitively, any element in $\monad(X)$ has a finite support. Moreover, 
if $\monad$ preserves weak pullbacks, then it also preserves finite intersection,
and one can define for $\mu \in \monad X$, the support 
of $\mu$ as:
\begin{align*}
\support{\mu} &\defeq \bigcap \{A \subseteq X \mid \mu \in \monad A\}.
\end{align*}
Since we deal with elements with countable support, we need to shift from 
finitary monads to $\aleph_1$-accessible monad. As a consequence, preservation 
of weak pullbacks is not enough to guarantee the existence of 
$\bigcap \{A \subseteq X \mid \mu \in \monad A\}$, as we now need $\monad$ 
to preserve \emph{countable} intersections.
\end{remark}


\subsection{Generic Effects}

Working with countable monads we can think of the support of a monadic 
object as its `value component'. What about its `effect component'? 
Here we show that such a component can be formally described using the 
notion of a \emph{generic effects} 
\cite{Plotkin/algebraic-operations-and-generic-effects/2003}. 
Achieving such a goal, however, requires to introduce some mathematical 
abstractions. As a consequence, we we first study a concrete 
example which will then generalise to arbitrary (countable) monads.

\begin{example}[Probabilistic computations]
\label{example:generic-effects-probabilistic-computations}
When working with (discrete) subdistributions, hence with the monad $\distribution$,
it is oftentimes convenient to represent subdistributions as 
syntactic objects called \emph{formal sums}. 
A formal sum (over a set $X$) is an expression of the form 
$\sum_{i \in I} p_i;x_i$, where $I$ is a countable set,
$p_i \in [0,1]$ and $x_i \in X$, and $\sum_i p_i \leq 1$. 

The notation $\sum_{i \in I} p_i;x_i$ is meant to recall the semantic 
counterpart of formal sums, namely subdistributions. However, we should keep 
in mind that formal sums are purely syntactical expressions. For instance, 
abusing a bit the notation, 
$\frac{1}{2}; x_0 + \frac{1}{2};x_0$ and $1;x_0$ are two distinct formal sums, 
although they both denote the Dirac distribution on $x_0$.
More generally, there is an interpretation function $\mathcal{I}$ mapping each formal sum 
$\sum_{i \in I} p_i;x_i$ to a subdistribution 
$\mu$ on $X$ defined as $\mu(x) \defeq \sum_{x_i = x} p_i$. Additionally, 
the map $\mathcal{I}$ is a \emph{surjection}, meaning that any 
subdistribution can be represented as a (non-unique) formal sum.

Examining a bit more carefully a formal sum 
$\sum_{i \in I} p_i;x_i$, we see that the latter consists of an $I$-indexed sequence 
$\lan p_i \ran_{i \in I}$ of elements in $[0,1]$ and an $I$-indexed sequence 
$\lan x_i \ran_{i \in I}$ of elements in $X$. 
Therefore, a formal sum is just a pair of sequences 
$(\lan p_i \ran_{i \in I}, \lan x_i \ran_{i \in I}) \in [0,1]^I \times X^I$ 
such that $\sum_i p_i \leq 1$. But the latter requirement means precisely that 
$\lan p_i \ran_{i \in I}$ is actually a subdistribution on $I$ (the one mapping 
$i$ to $p_i$). Therefore, we see that formal sums are just elements in 
$\distribution(I) \times X^I$.

Putting all these observations together, we see that for any $\mu \in \distribution(X)$, 
there exists a countable set $I$ and an element $F \in \distribution(I) \times X^I$ 
such that $\mathcal{I}(F) = \mu$. As a consequence, stipulating two formal sums 
$F_1, F_2 \in \distribution(I) \times X^I$ to be equal\footnote{
  Notice that $=_{\mathcal{I}}$ is the kernel of $\mathcal{I}$. 
}
(notation $F_1 =_{\mathcal{I}} F_2$)
if $\mathcal{I}(F_1) = \mathcal{I}(F_2)$, then we see that $\distribution(X)$ is 
isomorphic to the quotient set 
$(\bigcup_{I} \distribution(I) \times X^I)/=_{\mathcal{I}}$, where 
$I$ ranges over countable sets.
\end{example}

Summing up, Example~\ref{example:generic-effects-probabilistic-computations}
shows that any subdistribution $\mu \in \distribution(\values)$ can be decomposed 
as a pair $(\lan p_i \ran_{i \in I}, \lan \valone_i \ran_{i \in I})$, for some 
(countable) set $I$, where $\lan p_i \ran_{i \in I}$ is the `effect' part of 
$\mu$, and $\lan \valone_i \ran_{i \in I}$ is the value part of $\mu$. 
Even if the such a decomposition is not unique, we can give definitions and prove 
results on such decompositions and extend them to subdistributions by showing 
invariance with respect to $=_{\mathcal{I}}$, which is usually trivial. 

Can we generalize Example~\ref{example:generic-effects-probabilistic-computations} 
to arbitrary (countable) monads? 

First, let us observe that since the set $I$ in 
Example~\ref{example:generic-effects-probabilistic-computations} is countable, we can 
replace it with an enumeration of its elements. That is, we replace $I$ with 
sets $\makeset{n}$, where $n \in \mathbb{N}^{\infty} \defeq \mathbb{N} \cup \{\omega\}$ 
and\footnote{Counting from one rather than from zero simplifies the notation. 
We have also implicitly used this convention writing $\op(\termone_1, \hh, \termone_n)$ 
for $n$-ary algebraic operations.} 
$\makeset{n} \defeq \{1, \hh, n\}$ if $n \neq \omega$, and $\makeset{n} \defeq 
\mathbb{N}$, if $n = \omega$. 

Formally, we should work $\aleph_1$, 
the skeleton of the category of countable sets and all functions 
between them, instead of $\mathbb{N}^{\infty}$ (notice that $\aleph_1$ and
$\mathbb{N}^{\infty}$ have the same objects, but the former has `more arrows', so 
to speak). Nonetheless, thinking of $\mathbb{N}^{\infty}$ in place of 
$\aleph_1$ is perfectly fine for building intuitions. Abusing the notation, 
we write $n \in \aleph_1$ to state that $n$ is an object of $\aleph_1$.

\begin{theorem}[\cite{10.1007/978-3-642-99902-4_3,DBLP:journals/entcs/HylandP07,DBLP:journals/entcs/Power06a}]
\label{thm:representation}
For any set $X$ we have the isomorphism
    $$\monad X \cong \int^{n \in \aleph_1} \monad(\makeset{n}) \times X^n.$$
\end{theorem}
  
Let us decode Theorem~\ref{thm:representation}. First, 
$\int^{n \in \aleph_1} \monad(\makeset{n}) \times X^n$ is a \emph{coend} 
\cite{loregian2015coend}, an abstract notion that is not needed to achieve our goals. 
For us, $\int^{n \in \aleph_1} \monad(\makeset{n}) \times X^n$ is the quotient 
of $\coprod_{n \in \aleph_1} \monad(\makeset{n}) \times X^n$ for a suitable 
equivalence relation.

More precisely, we can translate Theorem~\ref{thm:representation} as follows:

\begin{theorem}
\label{thm:representation-2}
For any set $X$, all elements in $\monad(X)$ can be (non-uniquely) 
presented as elements in 
$$
\bigcup_{n \in \mathbb{N}^{\infty}} \monad(\makeset{n}) \times X^n
$$
Moreover, there is a \emph{surjective} map $\mathcal{I}$ such that for any 
$\mu \in \monad(X)$ there exists $n \in \mathbb{N}^{\infty}$ such that 
$\mu$ can be uniquely represented as an equivalence class modulo 
$=_{\mathcal{I}}$, the kernel of $\mathcal{I}$, 
of an element in $\monad(\makeset{n}) \times X^n$.
\end{theorem}

Replacing $\bigcup_{n \in \mathbb{N}^{\infty}} \monad(\makeset{n}) \times X^n$ 
with the more correct (yet morally equivalent) 
$\coprod_{n \in \aleph_1} \monad(\makeset{n}) \times X^n$, we achieve the correspondence
$$
\monad (X) \cong
\int^{n \in \aleph_1} \monad(\makeset{n}) \times X^n
=
{(\coprod_{n \in \aleph_1} \monad(\makeset{n}) \times X^n) / =_{\mathcal{I}}}
$$

Let us see how to prove Theorem~\ref{thm:representation-2}. 
In the following, we will oftentimes regard a 
sequence in $X^n$ as a function in $\makeset{n} \to X$.

\begin{proof}[Proof of Theorem~\ref{thm:representation-2}]
First, let us define the map $\mathcal{I}$. Given
$\gop \in \monad(\makeset{n})$ and $s: \makeset{n} \to X$ 
(where $n \in \aleph_1$), define 
$\mathcal{I}(\gop, s) \defeq \monad(s)(\gop)$.
Next we prove Surjectivity of $\mathcal{I}$. Let $\mu \in \monad(X)$. 
Since $\monad$ is countable, $\support{\mu}$ is countable, and thus 
isomorphic to $\makeset{n}$, for some object $n$ of $\aleph_1$.
any $\mu \in \monad(X)$ has a countable support. Let 
$f: \support{\mu} \to \makeset{n}$ be such a bijection. 
Then we present $\mu$ as 
$(\monad(f)(\mu), f^{-1})$. Since 
$\mathcal{I}(\monad(f)(\mu), f^{-1}) = \monad(f^{-1})(\monad(f)(\mu)) = 
\mu$ we are done. 
\end{proof}

We call elements in $\coprod_{n \in \aleph_1} \monad(\makeset{n}) \times X^n$ 
\emph{formal presentations}. Before giving examples of formal presentations 
we introduce a diagrammatic notation for them.

\section{A Diagrammatic Notation for Generic Effects}

Representing monadic elements as formal presentations has the major drawback 
of introducing bureaucracy in the treatment of indexes. 
Consider, for instance, a program of the form 
$\seq{\termone}{\termtwo}$. We try to define  
$\sem{\seq{\termone}{\termtwo}}$ using formal presentation.
First, we evaluate $\termone$ obtaining
$(\gop, \lan \valone_i \ran_{i \in \makeset{n}}) \in \monad(\makeset{n}) \times \values^n$, for some 
$n \in \aleph_1$. 
For any $i \in \makeset{n}$, we then evaluate 
$\subst{\termtwo}{\varone}{\valone_i}$, obtaining 
$(\goptwo_i, \lan \valtwo_{j_i} \ran_{j \in \makeset{m_i}})$, 
for some $\makeset{m_i} \in \aleph_1$. As a consequence, 
evaluating $\seq{\termone}{\termtwo}$ should give a generic effect 
$\gopthree  \in \monad(\makeset{l})$, for $l \defeq \sum_i m_i$, 
obtained by some kind of composition of $\gop$ with the 
$\goptwo_i$s together with the sequence 
$\lan y_{j_i}\ran_{j_i \in \makeset{l}}\ran$. 
The resulting expression is rather heavy to write.   
For this reason we introduce a diagrammatic notation for formal presentations.

\begin{definition}
\label{def:diagrmmatic-notation}
We represent an object 
$(\gop, \lan x_i \ran_{i \in \makeset{n}}) \in \monad(\makeset{n}) \times X^n
$ as a diagram of the form
$$
\eff{\gop}{n}{i}{x_i}
$$
Here $i$ ranges over elements in $\makeset{n}$ and to each $i$ it is associated the 
corresponding $x_i$. That is, the horizontal bar with subscript $i$ and 
target $x_i$ stands for the function $i \mapsto x_i$. 
\end{definition}

If $\makeset{n}$ is finite, we can modify Definition~\ref{def:diagrmmatic-notation} 
by extensionally listing all elements $x_1, \hh, x_n$. For instance, 

\begin{center}
\includegraphics[align=c,scale=1]{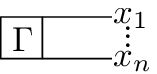}
\end{center} 
is the extensional version of 
$\eff{\gop}{n}{i}{x_i}$, for $n \in \mathbb{N}$.

\begin{example}
\begin{varenumerate}
\item Consider the maybe monad $\maybe$. We present an object
  $\mu \in \maybe(X)$ as a pair in 
  $\maybe(\makeset{n}) \times X^n$, 
  for some $n \in \mathbb{N}^{\infty}$. Since 
  $$
  \maybe(\makeset{n}) \times X^n = 
  (\makeset{n} + \{\divergence\}) \times X^n
  \cong 
  (\makeset{n} \times \Lambda^n) + (\Lambda^n \times \{\divergence\}),
  $$ 
  $\mu$ is (presented as) either a pair 
  $(k, \lan x_i \ran_{i \in \makeset{n}})$ or a pair 
  $(\divergence, \lan x_i \ran_{i \in \makeset{n}})$. 
  The former corresponds to the case of convergence to $x_k$, whereas the latter 
  to divergence.
  In particular, if $\mu$ 
  is the result of evaluating a $\lambda$-term, then we will actually have 
  $n = 1$ (if the term converges) or $n = 0$ (if the term diverges).
  \begin{itemize}
    \item If $n = 1$, we obtain pairs of the form $(1, \lan x \ran)$), which we write as
      $\effconvergence{x}$.
    \item If $n=0$, then we can only have the pair $(\bot,\lan \ran)$, where 
     $\lan\ran$ is the empty sequence. We write such a pair as $\boxed{\divergence}$.
  \end{itemize}
\item Consider the output monad $\Output$. We present an object
  $\mu \in \Output(X)$ as a pair in 
  $$
  \Output(\makeset{n}) \times X^n = 
  \mathcal{C}^{\infty} \times (\makeset{n} + \{\divergence\}) \times X^n,
  $$ 
  for some $n \in \mathbb{N}^{\infty}$. Therefore, $\mu$ is presented as 
  either a triple $(w, \divergence, \lan x_i \ran_{i \in \makeset{n}})$, 
  or as a triple $(w, k, \lan x_i \ran_{i \in \makeset{n}})$. 
  The former case means that we have divergence, and that the string $w$ is outputted, 
  whereas the latter case means that we converge to $x_k$, and that the string $w$ 
  is outputted. 
  As before, if $\mu$ is the result of evaluating a term, we will have 
  either $n = 1$ (if the term converges) or $n=0$ (if the term diverges). 
  \begin{enumerate}
    \item If $n=1$, we have triples of the form $(w, 1,x)$) which we write as 
       written as $\effnull{w}{x}$. 
    \item If $n=0$, the we can only have 
      triples of the form $(w,\bot,\lan\ran)$, which we write as 
      $\boxed{(w, \divergence)}$.
  \end{enumerate}
\item The case for the subdistribution monad goes exactly as in 
  Example~\ref{example:generic-effects-probabilistic-computations}. We can represent a 
  formal sum $(\lan p_i \ran_{i \in \makeset{n}}, \lan x_i \ran_{i \in \makeset{n}})$ as 
  $\eff{\lan x_i \ran_{i \in \makeset{n}}}{n}{i}{x_i}$. 
\item Replacing $([0,1], +)$ with $(\{0,1\}, \vee)$ in 
  Example~\ref{example:generic-effects-probabilistic-computations} we obtain 
  formal presentations for elements in $\powerset(X)$, which we may write 
  as diagrams  $\eff{\lan b_i \ran_{i \in \makeset{n}}}{n}{i}{x_i}$.
\end{varenumerate}
\end{example}

\subsection{A Calculus of Diagrams: Sequential Composition}

One of the strengths of monads (at least concerning their usage in the semantics of
programming languages) is that they naturally support 
the \emph{sequential} composition of effects. Do diagrams do the same?

Let us consider again $\sem{\seq{\termone}{\termtwo}}$. Suppose:
\begin{enumerate}
  \item $\sem{\termone} = \eff{\gop}{n}{i}{\valone_i}$;
  \item For any $i \in \makeset{n}$, $\sem{\subst{\termtwo}{\varone}{\valone_i}} 
   = \effm{\goptwo_i}{m_i}{j}{\valtwo_j}$.
\end{enumerate}

Can we find a diagram for $\sem{\seq{\termone}{\termtwo}}$? 
Since $\seq{\termone}{\termtwo}$ is nothing but a sequential composition, 
a natural proposal is to write 
$$
\efftwo{\hspace{0.05cm}\gop_{\phantom i}}{n}{i}{\goptwo_i}{m_i}{j}{\valtwo_j}
$$
for $\sem{\seq{\termone}{\termtwo}}$. But:
is this figure meaningful? That is, do we have 
a notion of composition for diagrams? 
We are going to answer such questions in the affirmative. 

\subsubsection{Sequential Composition}

There are at least two ways to define the sequential composition 
of two diagrams/formal presentations. The first one relies 
on the correspondence between generic effects and algebraic operations 
\cite{Plotkin/algebraic-operations-and-generic-effects/2003}.

\begin{proposition}
\label{prop:alg-op-gen-eff}
Any generic effect $\gop \in \monad(\makeset{n})$ corresponds to 
a $n$-ary algebraic operation $\gamma$, and any $n$-ary algebraic operation 
$\gamma$ corresponds to a generic effect 
$\Gamma \in \monad(\makeset{n})$. 
Moreover, if $(\Gamma, \lan x_i \ran_{i \in \makeset{n}})$ is a formal 
presentation of $\mu$, then $\mu = \gamma(\hh, \unit(x_i), \hh)$.
\end{proposition}

\begin{proof}
The bijection between algebraic operations and generic effects is given 
in \cite{Plotkin/algebraic-operations-and-generic-effects/2003}. 
Given $\gop \in \monad(\makeset{n})$ we define $\gamma_X: \monad(X)^n \to 
\monad(X)$ as $\gamma(\vect{x}) \defeq \kleisli{(\unit_X \circ \vect{x})}(\Gamma)$, where 
$\vect{x}: \makeset{n} \to X$. 
Vice versa, given $\gamma_X: \monad(X)^n \to \monad(X)$ we define 
$\gop \defeq \gamma_{\makeset{n}}(\unit_{\makeset{n}})$. 
Suppose now $(\Gamma, \lan x_i \ran_{i \in \makeset{n}})$ is a formal 
presentation of $\mu$. We write $\vect{x}$ for the sequence 
$\lan x_i \ran_{i \in \makeset{n}}$ regarded as a function $\makeset{n} \to X$. 
We have:
\begin{align*}
\mu &= \mathcal{I}(\Gamma, \lan x_i \ran_{i \in \makeset{n}}) 
\\
&= \kleisli{(\unit_X \circ \vect{x})}(\Gamma)
\\
&= \kleisli{(\unit_X \circ \vect{x})}(\gamma_{\makeset{n}}(\unit_{\makeset{n}}))
\\
&= \gamma_{\makeset{n}}(\kleisli{(\unit_X \circ \vect{x})} \circ \unit_{\makeset{n}})
\\
&= \gamma_{\makeset{n}}(\unit_X \circ \vect{x})
\end{align*}
where the penultimate equality follows by the defining 
identity law of algebraic operations 
(see Definition~\ref{def:algebraic-operation}).
\end{proof}

\begin{example}
\begin{varenumerate}
\item The generic effect corresponding to operation(s) 
  $\printsymbol{\mathtt{c}}: \Output(X) \to \Output(X)$ 
  is $(\mathtt{c},1)$
\item The generic effect corresponding to fair probabilistic choice $\oplus$ is the 
  distribution $\lan 0.5, 0.5\ran$ mapping $i \in \{1,2\}$ to $0.5$.
\end{varenumerate}
\end{example}

\begin{corollary}
\label{cor:composition-generic-effects-1}
There exists a notion of sequential composition for generic effects: given 
a generic effect $\gop \in \monad(\makeset{n})$ and a $\makeset{n}$-indexed family 
of generic effects
$\goptwo_i \in \monad(\makeset{m_i})$, there exists their sequential composition 
which is 
a generic effect in $\monad(\makeset{l})$, for 
$l \defeq \sum_{i \in \makeset{n}} m_i$.
\end{corollary}

\begin{proof}
Let us consider $\gop \in \monad(\makeset{n})$ and 
$\goptwo_i \in \monad(\makeset{m_i})$, for any $i \in \makeset{n}$.
The former gives the algebraic operation 
$\gamma: \monad(X)^{n} \to \monad(X)$. Similarly, the latter 
gives a family of algebraic operations 
$\delta_i: (\monad X)^{m_i} \to \monad(X)$. Taking products, 
we obtain
\[
\xymatrix{
 \monad(X)^{\sum_i m_i} \ar[r]^-{\cong}
 & \prod_{i \in \makeset{n}} \monad(X)^{m_i} 
 \ar[r]^-{\lan \delta_i \ran_{i \in \makeset{n}}} 
 & \monad(X)^n \ar[r]^-{\gamma} 
 & \monad(X)
}
\]
We then take advantage of Proposition~\ref{prop:alg-op-gen-eff} and 
define the composition of $\gop$ and $\goptwo_i$s 
as the generic effect associated to
the algebraic operation $\gamma \comp \lan \delta_i \ran_{i \in \makeset{n}}$.
\end{proof}

The drawback of Corollary~\ref{cor:composition-generic-effects-1} 
is that we would like to define 
the composition of generic effects without going through 
their associated algebraic 
operations. To achieve such a goal, we look at the \emph{countable Lawvere theory} 
\cite{citeulike:762448,DBLP:journals/entcs/HylandP07} induced by generic effects. 

First, observe that $\aleph_1$ has countable coproducts 
and, up-to equivalence, it is the free category with countable coproducts on 
$\makeset{1}$, 
where coproducts are given by standard sums 
(thus, for instance, for $n \in \mathbb{N}$ 
we have $n = \underbrace{1 + \cc + 1}_{n}$). By duality, $\aleph_1^{op}$ has 
countable products. 

\begin{definition}[\cite{DBLP:journals/entcs/Power06a}]
A \emph{countable Lawvere theory} 
is a small category $\mathcal{L}$ with countable products and a 
strict countable-product preserving identity-on-objects functor 
$I: \aleph_1^{op} \to \mathcal{L}$.
\end{definition}

In particular, objects of $\mathcal{L}$ are exactly those of 
$\aleph_1$, and every function between such objects gives a map 
in $\mathcal{L}$. A map in $\mathcal{L}(n,1)$ represents an 
\emph{operation} of arity $n$ (meaning that we also consider operations 
with arity $\omega$), whereas a map in $\mathcal{L}(n,m)$ represent $m$ 
operations of arity $n$. In particular, given operations 
$\op \in \mathcal{L}(n,m)$ and $\tau \in \mathcal{L}(m,l)$, their composition 
$\tau \comp \op \in \mathcal{L}(n,l)$ gives $l$ operations of arity $n$. 

\begin{proposition}
\label{prop:generic-effects-form-a-lawvere-theory}
Generic effects form a Lawvere theory, and thus we can define the composition 
of generic effects as composition on their Lawvere theory.
\end{proposition}

\begin{proof}
Let us begin showing that generic effects form a Lawvere theory. 
Let $Kl(\monad)_{\aleph_1}^{op}$ 
be the opposite of $Kl(\monad)$ restricted to objects of 
$\aleph_1$. Since $Kl(\monad)$ has countable coproducts, and the canonical 
functor $I: \set \to Kl(\monad)$ preserves them, restricting $I$ to 
$\aleph_1$ we obtain a countable-coproduct preserving identity-on-objects functor 
$I: \aleph_1 \to Kl(\monad)_{\aleph_1}$. As a consequence, $Kl(\monad)_{\aleph_1}$ 
is the opposite of a Lawvere theory, and thus
$Kl(\monad)_{\aleph_1}^{op}$ is a (countable) Lawvere theory. 
Notice that a generic effect $\gop \in \monad(\makeset{n})$ corresponds to a map in 
$Kl(\monad)_{\aleph_1}^{op}(n,1)$, i.e. as an operation in the Lawvere theory 
$Kl(\monad)_{\aleph_1}^{op}$.
We now define composition of generic effects.
Given generic a effect $\gop \in \monad(\makeset{n})$
and an $\makeset{n}$-indexed family of generic effects 
$\goptwo_i \in \monad(\makeset{m_i})$ (for any $i \in n$), i.e. arrows 
$\gop \in Kl(\monad)_{\aleph_1}^{op}(n,1)$ and 
$\goptwo_i \in Kl(\monad)_{\aleph_1}^{op}(p_i,1)$, we can define its 
composition relying on composition in $Kl(\monad)_{\aleph_1}^{op}(n,1)$ as follows. 
Since $\goptwo_i \in Kl(\monad)_{\aleph_1}^{op}(m_i,1)$ and 
$Kl(\monad)_{\aleph_1}$ has products, we have the map
$\lan \goptwo_i \ran_{i \in \makeset{n}} \in Kl(\monad)_{\aleph_1}^{op}(\sum_i m_i,n)$, 
and thus $\gop \comp \\lan \goptwo_i \ran_{i \in \makeset{n}} 
\in Kl(\monad)_{\aleph_1}^{op}(\sum_i m_i,1)$ is 
the desired generic effect. 
\end{proof}

Coming back to diagrams, given $\eff{\gop}{n}{i}{x_i}$ and 
$\effm{\goptwo_i}{m_i}{j}{y_j}$, for any $i \in \makeset{n}$, we 
write their composition as 
$$
\efftwo{\hspace{0.05cm}\gop_{\phantom i}}{n}{i}{\goptwo_i}{m_i}{j}{y_j}
$$
Such a notation gives us several advantages, as we are going to see. 
However, we first look at some concrete examples.

\begin{example}
\label{ex:diagram-composition}
\begin{varenumerate}
\item Consider the maybe monad and consider representations that can be 
  obtained as the result 
  of computations. Then sequential composition is defined by the following 
  laws:
   \begin{align*}
   \effnull{\divergence}{\mtermone} &\defeq \boxed{\divergence}
   \\
   \effnull{\downarrow}{\effnull{\downarrow}{\mtermone}} 
   &\defeq \effnull{\downarrow}{\mtermone}
 \end{align*} 
\item Consider the output monad. As before, we consider 
  representations that can be obtained as the result of computations. 
  Sequential composition is thus 
  defined:
  \begin{align*}
   \effnull{(w,\divergence)}{\mtermone} &\defeq \boxed{(w,\divergence)}
   \\
   \effnull{w}{\effnull{(u,\divergence)}{\mtermone}} 
   &\defeq \boxed{(wu,\divergence)}
   \\
   \effnull{w}{\effnull{u}{\mtermone}} 
   &\defeq \effnull{wu}{\mtermone}
 \end{align*} 
\item Consider the subdistribution monad, and denote by 
    $p \vect{q}$ scalar multiplication of $p$ with $\vect{q}$. 
    We then define sequential composition as:
    $$
    \efftwo{\vect{p}}{n}{i}{\vect{q}_i}{m_i}{j}{\mtermone_j}
    \defeq \effk{\vect{p} \cdot \lan \vect{q}_i \ran_{i \in \makeset{n}}}{l}{k}{\mtermone_k}
    $$
    where $l \defeq \sum_{i \in \makeset{n}} m_i$ and 
    $\vect{p} \cdot \lan \vect{q}_i \ran_{i \in \makeset{n}} \defeq 
    \lan p_i \vect{q_i} \ran_{i \in \makeset{n}}$ regarded as an $\makeset{l}$-indexed 
    sequence.
\end{varenumerate}
\end{example}

\subsection{A Calculus of Diagrams: Algebra}

Writing diagrams in sequence for sequential composition has important 
implications. First of all, the very act of writing diagrams in 
sequence for sequential composition presupposes a 
\emph{notion of equality} for diagrams. In fact, what we are doing is 
\emph{de facto} building an algebra of diagrams. Without much of a surprise, 
we say that two diagrams\footnote{We use letters 
$\mtermone, \mtermtwo, \mtermthree, \hh$ for diagrams} 
$\mtermone, \mtermtwo$ are equal (notation 
$\mtermone = \mtermtwo$) if and only if $\mtermone =_{\mathcal{I}} \mtermtwo$ 
is the map of Theorem~\ref{thm:representation-2}. 

It goes by itself that working with $=_{\mathcal{I}}$ directly may be 
quite heavy. For this reason, we now prove a collection of algebraic 
results that dispense us from working with $=_{\mathcal{I}}$. 
Such results concern sequential composition, monadic binding, and 
sequential composition.

\subsubsection{Algebra of Sequential Composition}

We immediately notice that composition is associative. 
This property is actually built-in the diagrammatic 
notation, as evident when writing the composition of three diagrams:
\begin{align}
\effthree{\hspace{0.05cm}\gop_{\phantom i}}{n}{i}{\goptwo_i}{m_i}{j}{\gopthree_j}{l_j}{k}\ 
x_{k}
\label{eq:three-diagrams}
\end{align}
Linguistically, this notation describes associativity of sequencing, usually 
expressed by program equivalences of the following form:
$$
\seq{\termone}{(\seqy{\termtwo}{\termthree})}
\equiv
\seqy{(\seq{\termone}{\termtwo})}{\termthree}.
$$

Diagram \eqref{eq:three-diagrams} also highlights another important feature of 
diagrams, namely the way they manage index dependencies. The `geography' of 
\eqref{eq:three-diagrams} 
shows that 
$i \in \makeset{n}$, $j \in \makeset{m_i}$, and $k \in \makeset{l_j}$. Moreover, 
by reading from the right to the left we recover index dependencies:
we see that since $k$ (being in $\makeset{l_j}$) depends on $j$ 
which depends (being in $\makeset{m_i}$) on $i$, $k$ depends on $i$ as well.

Second, there is a trivial generic effect $H \in \monad(\makeset{1})$ 
(read capital $\eta$) which behaves as a neutral element for composition. That is, 
for any $x \in X$, the diagram $\effunit{x}$ represent the computation trivially 
returning $x$ performing no effect (i.e. $\unit(x)$). 
We have the following laws:
\begin{align*}
\effunit{\mtermone} &= \mtermone
\\
\eff{\gop}{n}{i}{\effunit{x_i}} &= \eff{\gop}{n}{i}{x_i}
\end{align*} 

Linguistically, these laws correspond to the following program equivalences:
\begin{align*}
\seq{\valone}{\termone} &\equiv \subst{\termone}{\varone}{\valone}
\\
\seq{\termone}{\varone} &\equiv \termone
\end{align*}

\begin{example}
The object $\effunit{x}$ is $\effnull{\downarrow}{x}$ in the maybe monad $\maybe$, 
$\effnull{\varepsilon}{x}$ in the output monad $\Output$, 
and $\effnull{1}{x}$ in the 
subdistribution (resp. powerset) monad $\distribution$ (resp. $\powerset$).
\end{example}

Additionally, diagram equality is preserved by diagrams.

\begin{theorem}
\label{thm:diagram-binding}
Let $\mu \in \monad(X)$ be presented as 
$\eff{\gop}{n}{i}{x_i}$, and $f: X \to \monad(Y)$. 
Then $\mu \bind f$ is presented as 
$$
\eff{\gop}{n}{i}{f(x_i)}
$$
\end{theorem}

\begin{proof}
The proof relies on Proposition~\ref{prop:alg-op-gen-eff}, 
Corollary~\ref{cor:composition-generic-effects-1}, and the defining 
identity of algebraic operations of Definition~\ref{def:algebraic-operation}. 
Suppose $\mu$ is presented as 
$\eff{\gop}{n}{i}{x_i}$. By Proposition~\ref{prop:alg-op-gen-eff} we have an 
algebraic operation $\gamma: \monad(X)^n \to \monad(X)$ such that 
$\mu = \gamma(\hh, \unit(x_i), \hh)$. 
Suppose now 
$\effm{\goptwo_i}{m_i}{j}{y_j}$ is a presentation of 
$f(x_i)$, for any $i \in \makeset{n}$. In particular, by 
Proposition~\ref{prop:alg-op-gen-eff} we have algebraic operations
$\delta_i: \monad(Y)^{m_i} \to \monad(Y)$ such that 
$f(x_i) = \delta_i(\hh, \unit(y_j), \hh)$. 
Let 
$$
\mtermone \defeq 
\efftwo{\gop}{n}{i}{\goptwo_i}{m_i}{j}{y_j} 
= \eff{\gop}{n}{i}{f(x_i)}
$$
By Proposition~\ref{prop:alg-op-gen-eff} and 
Corollary~\ref{cor:composition-generic-effects-1}, $\mtermone$ is associated 
to the algebraic operation $\gamma \circ \lan \delta_i \ran_{i \in \makeset{n}}: 
\monad(Y)^{\sum_{i \in \makeset{n}} m_i} \to \monad(Y)$, and 
we have
$$
\mathcal{I}(\mtermone) = 
\gamma(\hh, \delta_i(\hh, \unit(y_j), \hh), \hh)
$$ 
We conclude the thesis as follows:
\begin{align*}
\mu \bind f &= \gamma(\hh, \unit(x_i), \hh) \bind f
\\
&=\gamma(\hh, \unit(x_i) \bind f, \hh)
\\
&=\gamma(\hh, f(x_i), \hh)
\\
&=\gamma(\hh, \delta_i(\hh, \unit(y_j), \hh), \hh)
\\
&= \mathcal{I}(\mtermone)
\end{align*}
where the second line follows by the defining 
identity of algebraic operations (see Definition~\ref{def:algebraic-operation}).
\end{proof}

An important consequence of Theorem~\ref{thm:diagram-binding} is the validity 
of the following law which makes most of the definitions given on diagrams 
invariant with respect to $=_{\mathcal{I}}$, and thus valid definitions on 
monadic elements\footnote{Notice that an analogous rule holds 
if we replace $f: X \to \monad(Y)$ with $f: X \to Y$.}.

\[
\infer{\eff{\gop}{n}{i}{f(x_i)} = \eff{\goptwo}{m}{j}{f(y_j)}
}
{\eff{\gop}{n}{i}{x_i} = \eff{\goptwo}{m}{j}{y_j}
&f: X \to \monad(Y)
}
\]

\begin{example}
Let us consider monadic terms, i.e. elements in $\monad(\Lambda)$. 
Recall that we have defined an evaluation map $\sem{-}: \Lambda \to \monad(\values)$. 
Define the monadic extension of $\sem{-}^{\monad}$ by:
$$
\left \llbracket \eff{\gop}{n}{i}{\termone_i} \right \rrbracket^{\monad} 
\defeq \eff{\gop}{n}{i}{\sem{\termone_i}}
$$
Then $\sem{-}^{\monad}$ is automatically well-defined, as by 
Theorem~\ref{thm:diagram-binding} $\sem{-}^{\monad}$ is invariant under diagram 
equality, and thus it is independent of the choice of representatives.
Notice also how the diagrammatic notation makes explicit how definitions 
and results `distribute' over effects. This is the very essence of algebraic 
effects, as stated in the defining 
identity of algebraic operations (Definition~\ref{def:algebraic-operation}). 
The diagrammatic notation extends this identity to any monadic expressions, 
as the latter are ultimately built using generic effects, which bijectively 
correspond to algebraic operations. 
\end{example}

\subsubsection{Algebraic Operations}

Thanks to Proposition~\ref{prop:alg-op-gen-eff}, diagrams are also 
a natural way to write algebraic operations.
In general, if we present objects $\mu_i \in \monad(X)$ as $\mtermone_i$, then 
we write 
\begin{center}
\includegraphics[align=c,scale=0.9]{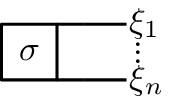} 
$\quad$ or $\quad$ $\eff{\phantom{|}\op\phantom{|}}{n}{i}{\mtermone_i}$
\end{center} 
for the presentation 
of $\op(\mu_1, \hh, \mu_n)$ (that is, we use the notation $\op$ both for 
the algebraic operation and the generic effects associated to it).

Notice that thanks to Theorem~\ref{thm:diagram-binding} both 
$\op(\mu_1, \hh, \mu_n) \bind f$ and $\op(\mu_1 \bind f, \hh, \mu_n \bind f)$ 
are presented as 
$$
\eff{\op}{n}{i}{f(\mtermone_i)},
$$ where $\mtermone_i$ is a presentation 
of $\mu_i$, this way encoding the defining identity of algebraic operations 
(Definition~\ref{def:algebraic-operation}) in diagrams. 
Linguistically, this corresponds to the program transformation 
$$
\seq{\op(\termone_1,\hh, \termone_n)}{\termtwo} 
\equiv
\op(\hh, \seq{\termone_i}{\termtwo},\hh).
$$
Having seen the basic algebra of diagrams, we move to the study of 
their order-theoretic properties.

\subsection{A Calculus of Diagrams: Order}

So far we have focused on algebraic properties of diagrams. However, 
working with $\signature$-continuous monads, we can extend the 
order $\cpoleq$ on diagrams by stipulating $\mtermone \cpoleq \mtermtwo$ 
if and only if $\mathcal{I}(\mtermone) \cpoleq \mathcal{I}(\mtermtwo)$.
This way, we see that diagrams
enjoy plesant order-theoretic properties. Before studying such properties, however,
it is useful to relate the elements $\lan x_i \ran_{i \in \makeset{n}}$ in
a diagram $\eff{\gop}{n}{i}{x_i}$ presenting $\mu$ with the support of 
$\mu$.

\paragraph{On Support}

Given an element $\mu \in \monad(X)$ presented as a diagram 
$\eff{\gop}{n}{i}{x_i}$, we see that the set $\{x_i \mid i \in n\}$ 
is a superset of $\support{\mu}$. For instance, 
the Dirac distribution on $x$ (i.e. $\unit(x)$) is presented as 
$\effunit{x}$ (i.e. as the formal sum $1;x$). However, we can also represent 
it as the formal sum $1;x + 0;y$. 

More generally, given $\gop \in \monad (\makeset{n})$ and 
$\makeset{n} \xhookrightarrow{\ \iota\ } \makeset{m}$, we also have the injection 
$\monad(\makeset{n}) \xhookrightarrow{\ \monad(\iota)\ } \monad(\makeset{m})$. 
For instance, taking
$\monad = \distribution$ and $\lan p_i \ran_{i \in \makeset{n}} 
\in \distribution(\makeset{n})$, 
the subdistribution $\distribution (\iota)\lan p_i \ran_{i \in \makeset{n}}$ 
maps $i$ to $p_i$ if $i \in \makeset{n}$, and to $0$ otherwise 
(i.e. if $i \in \makeset{m} \setminus \makeset{n}$).

Formally, given a countable set $X$, 
any formal presentation $(\gop, \vect{x}) \in \monad(\makeset{s}) \times X^n$ 
can be extended to a formal presentation $(\goptwo, \vect{y})$ 
in $\monad(\makeset{m}) \times X^m$, for any 
$\makeset{n} \xhookrightarrow{\ \iota\ } \makeset{m}$.

\begin{lemma}
Let $X$ be a countable set. Then any pair 
$(\gop, \vect{x}) \in \monad(\makeset{n}) \times X^n$ 
can be extended to a formal presentation $(\goptwo, \vect{y})$ 
in $\monad(\makeset{m}) \times X^m$, for any 
$\makeset{n} \xhookrightarrow{\ \iota\ } \makeset{m}$
\end{lemma}

\begin{proof}
Since $\makeset{n} \xhookrightarrow{\ \iota\ } \makeset{m}$, then
$\monad(\makeset{n}) \xhookrightarrow{\ \monad(\iota)\ } \monad(\makeset{m})$. Take 
$\goptwo \defeq \monad (\iota)(\gop)$ and $\vect{y}$ to be any map 
extending $\vect{x}$ according to the following diagram (such maps indeed exists):
$$
\xymatrix{\makeset{m} \ar[r]^{\vect{y}} & X
\\
\makeset{n} \ar@{^{(}->}[u]^{\iota} \ar[ru]_{\vect{x}}}
$$
Indeed $(\gop, \vect{x}) = (\goptwo, \vect{y})$.
\end{proof}

For instance, we have seen that the trivial generic effect $H \in \monad(\makeset{1})$ 
corresponds to the map $\unit_1: \makeset{1} \to \monad(\makeset{1})$. However, 
for any $n$, there is a map $\unit_{\makeset{n}}: \makeset{n} \to \monad(\makeset{n})$ 
such that, for any 
$i \in \makeset{n}$, $\unit(i) = \monad(\iota)(H)$, 
for $\makeset{1} \xhookrightarrow{\ \iota\ } \makeset{n}$ 
sending $1 \in \makeset{1}$ to $i$. As a consequence, we see that we can indeed think of 
$H$ as the `real' trivial effect, and regard the others as its extension 
to larger supports. For instance, we can regard $1;x$ as \emph{the} presentation of 
the Dirac distribution on $x$, and formal sums such as 
$1;x + 0; y + 0;z$ as extensions of $1;x$.

\paragraph{Order-theoretic Properties}

We now analyze the order-theoretic properties of diagrams. 
For technical reason, we need to require $\monad(\makeset{1})$ to have 
at least two elements\footnote{
This property ensures the possibility to embed the Boolean algebra 
(which is isomorphic to) $\makeset{2}$ to $\makeset{1}$. Notice also 
that if $\bot = \unit(1)$, then $\monad(\makeset{1}) \cong \makeset{1}$.
It is a straightforward exercise to verify that all monads mentioned so 
far satisfy this condition.}, i.e. $\unit(1) \neq \bot$.
First, 
we observe that there is a bottom effect $\bot \in \monad(\makeset{0})$, 
which we write as $\boteff$. It is easy to see that such element 
is $\boxed{\divergence}$ in the maybe monad, $\boxed{(\varepsilon, \divergence)}$
(where $\varepsilon$ is the empty string)
in the output monad, and $\boxed{0}$ in the subdistribution monad. 

Actually, for any $n$ there is a bottom effect $\bot_n \in \monad(\makeset{n})$
such that $\eff{\bot_n}{n}{i}{x_i} \cpoleq \eff{\gop}{n}{i}{y_i}$. 
This is a bit unsatisfactory, as we might expect 
the `real' bottom effect to be $\boteff$. 
That is actually the case.

In fact, since $\monad$ is $\signature$-continuous, 
$\monad(f)$ is strict, for any $f: X \to Y$. Therefore, for any $n$,
the map $\monad(\makeset{0} \xhookrightarrow{\ \iota\ } \makeset{n})$ is strict, 
meaning that $\bot_n = \monad (\iota)(\bot)$. That means that we essentially 
have a unique bottom effect, viz. $\boteff$, and we can 
regard any $\bot_n$ as its extension to larger supports.
In fact, we have the equality
$$
\eff{\bot_n}{n}{i}{\mtermone_i} = \boteff
$$
as well as the inequality
$$
\boteff \cpoleq \mtermone
$$
For this reason we write $\bot$ in place of $\bot_n$. 

We may also ask whether the `dual' equality holds, i.e. if 
$$
\eff{\gop}{n}{i}{\bot} = \bot
$$
This is not the case. In fact, there are algebraic operations such as $\printsymbol{-}$ 
that are not strict. However, if effects are \emph{commutative}, then 
such an equality holds.

\begin{definition}
\label{def:commutative-monad}
We say that a monad is \emph{commutative} if
$$
\eff{\gop}{n}{i}{\eff{\goptwo}{m}{j}{x_{i,j}}} 
=
\eff{\goptwo}{m}{j}{\eff{\gop}{n}{i}{x_{i,j}}}
$$
Notice that $m$ is independent of $i$, and 
$n$ of $j$.
\end{definition} 

Please observe that Definition~\ref{def:commutative-monad} is rather  
different from standard definitions of commutative monads one meets in the literature.
Indeed, one of 
the advantages of our notation is to allow for a simple, operational 
definition of commutativity of monads. Accordingly, we easily notice that 
the commutativity equation in Definition~\ref{def:commutative-monad} 
is just the semantical counterpart of the following program equivalence:
$$
\seq{\termone}{(\seqy{\termtwo}{\termthree})}
\equiv
\seqy{\termtwo}{(\seq{\termone}{\termthree})}
$$
where $\varone \in FV(\termtwo)$ and $\vartwo \not \in FV(\termone)$.

\begin{proposition}
If $\monad$ is commutative, then 
$$
\eff{\gop}{n}{i}{\bot} = \bot
$$
\end{proposition}
\begin{proof}
Calculate
\begin{align*}
  \eff{\gop}{n}{i}{\bot} &= \eff{\gop}{n}{i}{\eff{\bot_n}{n}{i}{i}}
  \\
  &= \eff{\bot_n}{n}{i}{\eff{\gop}{n}{i}{i}}
  \\
  &=\boteff
\end{align*}
\end{proof}

Having clarified the role of the bottom effect, let us now 
move to monotonicity laws.

\begin{proposition}
\label{prop:monotonicity-laws}
  The following monotonicity laws hold:
  \[
  \infer{\eff{\gop}{n}{i}{f(x_i)} \cpoleq \eff{\goptwo}{m}{j}{f(y_j)}}
  {\eff{\gop}{n}{i}{x_i} \cpoleq \eff{\goptwo}{m}{j}{y_j} 
  & f: X \to \monad(Y)}
  \]
  \[  
  \infer{\eff{\gop}{n}{i}{f(x_i)} \cpoleq \eff{\gop}{n}{i}{g(x_i)}}
  {f,g: X \to \monad(Y)
  & f \cpoleq g
  }
  \]
\end{proposition}

\begin{proof}
We prove the first rule (the second is similar).
Let 
$\mtermone \defeq \eff{\gop}{n}{i}{f(x_i)}$, 
$\mtermtwo \defeq \eff{\goptwo}{m}{j}{f(y_j)}$, 
$\gamma$ the algebraic operation corresponding to $\gop$, 
and $\delta$ the one corresponding to $\Delta$. Notice that 
by hypothesis we know $\gamma(\hh, \unit(x_i),\hh) \cpoleq
\delta(\hh, \unit(x_i),\hh)$, and recall that $\kleisli{f}$ 
is monotone.
We have
$\mathcal{I}(\mtermone) \cpoleq \mathcal{I}(\mtermtwo)$. 
For:
\begin{align*}
\mathcal{I}(\mtermone) 
&= 
\kleisli{f}(\gamma(\hh, \unit(x_i),\hh))
\\
&\cpoleq 
\kleisli{f}(\delta(\hh, \unit(x_i),\hh))
\\
&= 
\mathcal{I}(\mtermtwo).
\end{align*}
\end{proof}

\begin{corollary}
The following monotonicity law hold:
  \[
  \infer{\eff{\gop}{n}{i}{\mtermone_i} \cpoleq \eff{\gop}{n}{i}{\mtermtwo_i}}
  {\forall i \in \makeset{n}.\ \mtermone_i \cpoleq \mtermtwo_i}
  \]
\end{corollary}

\begin{proof}
Define $f,g: \makeset{n} \to \monad(X)$ by $f(i) \defeq \mtermone_i$ and 
$g(i) \defeq \mtermtwo_j$. Clearly, we have $f \cpoleq g$. Moreover, since 
$\eff{\gop}{n}{i}{\mtermone_i} = \eff{\gop}{n}{i}{f(i)}$
and 
$\eff{\gop}{n}{i}{\mtermtwo_i} \cpoleq \eff{\gop}{n}{i}{g(i)}$
we conclude the thesis by Proposition~\ref{prop:monotonicity-laws}.
\end{proof}

\section{Conclusion}

We have introduced a diagrammatic notation and calculus for 
countable monads with algebraic operations. We have shown by means of 
examples and general results some of advantages of our notation. 
Additionally, the authors are currently using this notation to prove new,
nontrivial theorems on calculi with algebraic effects whose proofs turned out 
to be extremely heavy using the standard, linear notation.
We leave as a future work the investigation of further applications of 
such a notation.

\bibliography{main}

\end{document}